\documentstyle[12pt]{article}
\newcommand{\be}{\begin{equation}}
\newcommand{\ee}{\end{equation}}
\begin{document}

\title{The Attractive Nonlinear Delta-Function Potential}
\author{{\bf M. I.  Molina and C. A. Bustamante}
\vspace{1 cm}
\and
\and
Facultad de Ciencias, Departamento de F\'{\i}sica, Universidad de Chile\\
Casilla 653, Las Palmeras 3425, Santiago, Chile.}
\date{}
\maketitle
\baselineskip 24 pt
\vspace{2in}
\newpage
\centerline{\bf Abstract}
\vspace{1.5cm}

\noindent
We solve the continuous one-dimensional Schr\"{o}dinger equation for the
case of an inverted {\em nonlinear} delta--function potential located
at the origin, obtaining the bound state in closed form as a function of the
nonlinear exponent. The bound state probability profile decays
exponentially away from the origin, with a
profile width that increases monotonically with the nonlinear
exponent, becoming an almost completely extended state when this 
approaches two. At an exponent value of two, the bound
state suffers a discontinuous change to a delta--like profile. Further
increase of the exponent increases again the width of the probability
profile, although the bound state is proven to be stable only for
exponents below two. The transmission of plane waves across the nonlinear
delta potential increases monotonically with the nonlinearity
exponent and is insensitive to the sign of its opacity.

\newpage

\noindent
The delta-function potential $\delta(x - x_{0})$ has become a familiar sight
in the landscape of most elementary courses on quantum mechanics, where
it serves to illustrate the basic techniques in simple form. As a physical
model, it has been used to represent a localized potential whose energy scale
is greater than any other in the problem at hand and whose spatial extension is
smaller than other relevant length scales of the problem.
Arrays of delta-function potentials have been
used to illustrate Bloch's theorem in solid state physics and also in optics,
where in the scalar approximation, wave propagation in a periodic medium
resembles the dynamics of an electron in a crystal lattice.
It is well known that the single ``inverted'' delta-function
potential $-\Omega\ \delta(x-x_{0})$ possesses one exponentially
localized bound state for all values of the opacity parameter $\Omega$.
Its existence and stability
has been tested against the effects of different boundary
conditions$^{1}$ and symmetry-breaking perturbations$^{2}$.
In addition, the inverted delta potential 
has been used as a semi-permeable barrier to examine resonance
phenomena in scattering theory$^{3}$,
among others. Other interesting applications of the delta-function
potential concept
are found in Ref.4.

\noindent
In this work we examine the problem of finding the bound state
and the transmission coefficient of plane waves across an inverted
{\bf nonlinear} delta--function potential,
described by the so-called Nonlinear Schr\"{o}dinger (NLS) equation:
\be
-{\hbar^{2}\over{2 m}} \phi''(x) -
{\hbar^{2}\over{2 m}}\ \Omega\ \delta(x)\ |\phi(x)|^{\alpha}\ \phi(x) =
E\ \phi(x),\label{eq:1}
\ee
where $\Omega > 0$ is the opacity coefficient and $\alpha$ is the
nonlinearity exponent.
For $\alpha = 0$ we recover the familiar problem of the linear
``inverted'' delta-potential
which possesses an exponentially decaying bound state profile for
any opacity strength:
$\phi(x) = \sqrt{\Omega/2} \exp(-(\Omega/2)\ |x| )$. The rate of
decay in space is determined by the
localization length $2/\Omega$ which increases (decreases) as the
opacity decreases (increases). 

\noindent
It might seem odd at first
to see a {\em nonlinear--looking} Schr\"{o}dinger equation like
(\ref{eq:1}), since we know that quantum mechanics is linear.
Therefore, all physical systems should be described by coupled
sets of linear equations. However we oftentimes
can only concentrate on a few ``relevant'' degrees of freedom,
 making suitable approximations to deal with
the rest. At times, the price to pay for this reduction is
the appearance of nonlinear
evolution equations for the variables of
interest, such as Eq. (\ref{eq:1}).

\noindent
For instance, in atomic physics, a
well-known approximation when dealing with
multi-electron atoms is the self-consistent field approximation
(Hartree-Fock). In this case each  electron is described
by a single-particle wave function that solves a Schr\"{o}dinger-like equation.
The potential appearing in this
equation is that generated by the average motion of all the other
electrons, and so depends on their
single-particle wave functions. This results in a set of nonlinear
eigenvalue equations$^{5}$. A more recent
application of the mean-field ideas to a weakly interacting Bose
condensate can be found in Ref.[6].

\noindent
For $\alpha = 2$, Eq.(\ref{eq:1}) could model the problem of
an electron propagating in a
one--dimensional linear medium which contains a vibrational ``impurity''
at the origin that can couple strongly
to the electron. In the
approximation, where one considers the vibrations completely ``enslaved''
to the electron, one obtains Eq.(\ref{eq:1}) as the
effective equation for the electron.

\noindent
A closely related equation,
given by
the {\em discrete} version of (\ref{eq:1}) is known as the discrete
nonlinear Schr\"{o}dinger (DNLS)
equation.  It was introduced in its time--independent form,
in the late fifties by Holstein in his studies of the
polaron problem$^{7}$ in
condensed matter physics. The DNLS equation was derived
in a fully time--dependent
form by Davydov in his studies of energy transfer
in proteins and other biological
materials$^{8}$. In the continuum limit
the time-dependent DNLS equation reduces to the time-dependent 
NLS equation, which supports soliton solutions. Therefore, a
soliton--based energy transport appears as an attractive
candidate mechanism for energy transport in biomolecules. A recent
review of the status of Davydov's proposal can be found
in Ref.[9].
The time-dependent DNLS equation can also be viewed as the
evolution equation for a Hamiltonian system of classical
anharmonic oscillators$^{10}$.

\noindent
An important application of the continuous model (\ref{eq:1})
is that of a wave propagating in a one--dimensional linear
medium which contains a narrow strip of nonlinear (general
Kerr-type) material$^{11}$. This nonlinear strip is assumed
to be much smaller than the typical wavelength.
In fact, periodic and quasiperiodic arrays of nonlinear strips 
have been considered by a straightforward generalization of
Eq.(\ref{eq:1}) in order to model wave propagation in
some nonlinear superlattices$^{12}$. 

\noindent
{\bf Bound State ($E = E^{(b)} < 0$):}\ \  Our system consists of a single,
infinitely localized potential well in a continuous infinite line and
therefore,  lacks any natural length scale.
If the delta potential were confined between two infinite walls,
the distance between the walls would provide a length scale. If,
instead of a continuous line, the potential were defined on a
discrete lattice, its lattice constant would define a natural
length scale. Also, if instead of one delta potential, we had
at least two of them, their mutual distance
would constitute a natural length scale for the system.

\noindent
In our case we have none of these. Thus,  $\Omega$ serves only
to define the unit of distance (as it does in the linear case).
It is possible to get rid of $\Omega$ formally as follows:
From Eq.(\ref{eq:1}) we see that $\Omega$ must have units of
$[\mbox{distance}]^{(\alpha/2) - 1}$, which suggests the definition
of a dimensionless distance $u$ as
$u = x/L$, with
$L\equiv \Omega^{2/(\alpha -2)}$.
In terms of $u$ and $\phi(u)\equiv (1/\sqrt{L})\ \phi(x)$,
Eq.(\ref{eq:1}) can be recast in a dimensionless form:
\be
\phi''(u) - k^{2}\ \phi(u) = -\delta(u)\ |\phi(u)|^{\alpha}\ \phi(u),\label{eq:5}
\ee
with $k^{2} = -2 m L^{2} E^{(b)}/{\hbar^{2}}$ and $\phi''(u) = (d^{2}/du^{2}) \phi(u)$.
The opacity $\Omega$ has now disappeared from view since it only
determines the unit of distance. We try:
\be
\phi(u) = \left\{ \begin{array}{ll}
	A\ \exp(k\ u)	& 	\mbox{$u < 0$}\\
	B\ \exp(-k\ u)	&	\mbox{$u > 0$}
	\end{array}	\right.
\ee
Using the continuity of $\phi(u)$ and the discontinuity of $\phi'(u)$ at $u = 0$, one obtains
$A = B$ and $k = (1/2)\ |A|^{\alpha}$. Finally, use of the normalization
condition $ 1 = \int_{-\infty}^{\infty} |\phi(u)|^{2}\ du$, leads to 
\be
\phi(u) = \left( {1\over{2}} \right)^{1/(2 - \alpha)}\ \exp\left[-
\left( {1\over{2}} \right)^{2/(2 - \alpha)}\ |u| \right] \label{eq:7}
\ee
with a dimensionless bound state energy
\be
E^{(b)} = - \left( {1\over{2}} \right)^{4/(2-\alpha)}.\label{eq:77}
\ee

\noindent
As in the linear ($\alpha =0$) case, the bound state profile is
exponentially decreasing away from the delta
potential with localization length $2^{2/(2-\alpha)}$.
As $\alpha$ increases from zero, the probability
profile widens and the bound state energy decreases in magnitude.
At $\alpha = 2^{-}$, the state is completely
extended all over the real axis and the bound state energy is
vanishingly small. At $\alpha=2^{+}$, the bound
state becomes infinitely localized,
with a delta--like probability profile and with an infinite bound
state energy. Further increase in the nonlinear
exponent leads to a widening of the probability profile and to a
corresponding reduction
in the magnitude of the bound state energy.

\noindent
Figure 1 shows the
amplitude, or the inverse square
probability profile width,  of the bound state as a function of
the nonlinearity exponent.
However, at this point, an important observation
is in order. The total {\em energy} of the system does not coincide
with the bound state energy. In order to see this, we must consider the
full time-dependent nonlinear Schr\"{o}dinger equation that gives rise
to Eq. (\ref{eq:5}). By using $\tau\equiv t/T$ as a
dimensionless time variable, with $T\equiv (\hbar/2 m L^{2})^{-1}$,
we have
\be
i\ {d\ \psi(u, \tau)\over{d\ \tau}} = - {d^{2}\ \psi(u, \tau)\over{d\ \tau^{2}}}  -
\delta(u)\ |\psi(u, \tau)|^{\alpha}\ \psi(u, \tau). \label{eq:total}
\ee
In other words, we have $i (d/d \tau) \ \psi(u, \tau) = H\ \psi(u, \tau)$,
where the Hamiltonian operator can be decomposed as
$H = H_{0} + O_{NL}$,  where $H_{0} = p^{2}$, with $p = i\ (d/d u)$
and $O_{NL} = -\delta(u)\ |\psi(u,\tau)|^{\alpha}$ as the nonlinear part.
We see that $H$ depends on time explicitly, through the time dependence
of $O_{NL}$:
\be
{\partial H\over{\partial \tau}} = {\partial O_{NL}\over{\partial \tau}}\label{eq:10}
\ee
This implies that $\langle H \rangle$ is no longer a constant of the motion:
\be
{d \langle H \rangle \over{d \tau}} = i \langle \ [H, H]\ \rangle +
\left\langle {\partial H \over{\partial \tau}}\right\rangle =
\left\langle {\partial H \over{\partial \tau}}\right\rangle  \neq 0. \label{eq:11}
\ee
For the nonlinear part, we have
\be
{d \langle O_{NL} \rangle \over{d \tau}} = i \langle \ [H, O_{NL}]\  \rangle +
\left\langle {\partial O_{NL} \over{\partial \tau}}\right\rangle.
\ee
But,
\be
\left\langle {\partial O_{NL} \over{\partial \tau}}\right\rangle =
\int du |\psi(u,\tau)|^{2} (\partial O_{NL}/\partial \tau).\label{eq:d-}
\ee
By expressing $O_{NL}$ in terms of $\psi(u,
\tau)$ and using Eq. (\ref{eq:total}), 
we can recast (\ref{eq:d-}) as
\be
\left\langle {\partial O_{NL} \over{\partial t}}\right\rangle =
{i \alpha\over{2}}\ \langle [H, O_{NL}] \rangle,\label{eq:d}
\ee
which means
\be
{d \langle O_{NL} \rangle \over{d \tau}} = i\ \left( 1 +
{\alpha\over{2}} \right) \langle \ [H, O_{NL}]\  \rangle.\label{eq:dd}
\ee
By comparing Eqs. (\ref{eq:d}) and (\ref{eq:dd}), we conclude
\be
\left\langle{\partial O_{NL} \over{\partial \tau}}\right\rangle =
\left( {\alpha\over{\alpha + 2}}\right) \ {d\over{d \tau}} \langle O_{NL} \rangle.
\ee
Finally, by inserting this back into Eq.(\ref{eq:10}),
Eq.(\ref{eq:11}) becomes:
\be
{d \over{d \tau}} \langle H \rangle =
\left( {\alpha\over{\alpha + 2}}\right)\ {d \over{d \tau}} \langle O_{NL} \rangle,
\ee
which implies
\be
{d \over{d \tau}}\left\langle H - \left( {\alpha\over{\alpha + 2}}\right) O_{NL} \right\rangle = 0.
\ee
Therefore, the true energy operator
for our system is
$H_{t} \equiv H - (\alpha/\alpha +2)\ O_{NL}$.
For a stationary-state, $\psi(u,\tau) = \phi(u) \exp(-i E^{(b)} t/\hbar)$,
the total dimensionless energy is then
\be
E_{t} = E^{(b)} -
\left( {\alpha\over{\alpha + 2}} \right) ( -|\phi(0)|^{\alpha +2} ) =
- \left( {1\over{2}} \right)^{4/(2 - \alpha)}\ \left( {2 - \alpha\over{2 + \alpha}} \right).
\ee
Thus, for $\alpha <2$ the total energy is negative and the eigenstate is a stable
localized state. On the contrary, when $\alpha >2$, the total energy is positive
and the eigenstate is localized but possibly unstable, which means that any weak
`perturbation' could make it disappear into the continuum.  This explains the `stable'
and `unstable' labelling in Fig.1. Only for $\alpha = 0$, i.e., the linear case,
both the total energy and the energy eigenvalue coincide.
Figure 2 shows some probability profiles for several
different values of the nonlinear exponent that give rise to true (stable) bound states.
This distinction between the eigenenergy and the total energy must always be kept in mind
when dealing with effectively nonlinear systems. 

\noindent
{\bf Transmission of plane waves ($E > 0$):}\ \ We now cast Eq.(\ref{eq:1}) as
\be
\psi''(x) + k^{2}\ \psi(x) = -\Omega\ \delta(x)\ |\psi(x)|^{\alpha}\ \psi(x)\label{eq:scat}
\ee
where $k^{2} = 2 m E/{\hbar^{2}}$ is the electron wavevector.
Unlike the bound state problem, we now 
have $1/k$ as a natural length scale and can therefore consider $\Omega$ as
a {\em bona fide} opacity coefficient. 
The problem looks similar to the usual single delta-barrier problem,
with the exception of the nonlinear term $| \psi |^{\alpha} $ that
modulates the strength of the barrier opacity, depending on how much electronic
probability is sitting on the barrier. We will examine the dependence of the
transmission coefficient on $\Omega$ and $\alpha$.

\noindent
Since we are interested in plane wave transmission, we set
\be
\psi(x) = \left \{ \begin{array}{ll}
	   R_{0} \exp(i k x) + R \exp(-i k x)	&\ \ \ \ \ \ \mbox{$x < 0$}\\
	   T \exp(i k x)			&\ \ \ \ \ \ \mbox{$x > 0$}
	   \end{array}
	   \right.
\ee
From the continuity of $\psi(x)$ and discontinuity of $\psi'(x)$ at $x = 0$, we obtain
\begin{eqnarray}
T 						  & = & R_{0} + R\\
i k T 			  & = & i k (R_{0} - R) - \Omega\ |T|^{\alpha}\ T.
\end{eqnarray}

From here, one obtains $T = 2\ R_{0}/( 2 - (i \Omega/k)\ |T|^{\alpha})$.
Defining the transmission coefficient as $ t \equiv | T |^{2} / | R_{0} |^{2} $, we obtain
the following equation for the transmission coefficient:
\be
t ={4\over{4 + (\Omega^{*}/k)^{2}\ t^{\alpha}}}\label{eq:t}
\ee
where $\Omega^{*} \equiv \Omega |R_{0}|^{\alpha}$ is the ``effective''
opacity. We note that (\ref{eq:t}) is invariant under a sign change
in $\Omega$. In other words, both the
``upright'' and the ``inverted'' delta potentials possess identical transmissivities.

For arbitrary $\alpha$, Eq. (\ref{eq:t}) is a nonlinear equation for $t$ and must be solved
numerically. There are, however, four exactly solvable cases, three of which can
be described shortly:
\begin{enumerate}
\item
$\alpha = 0$ (linear case):\ \ From (\ref{eq:t}) we immediately obtain the well--known result
\be
t = {1\over{1 + (\Omega/2 k)^{2}}}
\ee
\item
$\alpha = 1$:\ \ Now Eq. (\ref{eq:1}) can be recast as the quadratic equation 
$(\Omega^{*}/ 2k)^{2}\ t^{2} + t -1 =0$, with physical solution
\be
t = 2 \left({k\over{\Omega^{*}}}\right)^{2}\ \left[ -1 + \sqrt{1 + \left({\Omega^{*}\over{k}} \right)^{2}} \right]
\ee
\item
$\alpha = 2$:\ \ Now we deal with a cubic equation for $t$: $(\Omega^{*}/2 k)^{2} t^{3} + t -1 =0$.
Its physical solution is
\be
t = (2/9)^{1/3}\ (1/|\Omega^{*}|)\ A(k,\Omega^{*}) - (32/3)^{1/3}\ (k^{2}/|\Omega^{*}|)\ A(k,\Omega^{*})^{-1/3}
\ee 
where $A(k,\Omega^{*}) = 9 k^{2} |\Omega^{*}| + \sqrt{3 (16 k^{6} + 27 k^{4} {\Omega^{*}}^{2})}$
\end{enumerate}
The case $\alpha =3$ is exactly solvable in principle, but it leads to a cumbersome 
expression for $t$ that is not particularly illuminating.

If we recast the general equation for $t$ as $t\ ( 1 + (\Omega^{*}/2 k)^{2}\ t^{\alpha}) = 1$, a bit of
simple analysis will convince the reader that the left hand side is always a monotonically
increasing function of $t$ for $\alpha > 0$. Therefore, there is always only one solution in  the
interval $0 \leq t \leq 1$. Figure 3 shows the transmission coefficient $t$ as a function of
$k/\Omega^{*}$ and several different nonlinearity exponents $\alpha$. Unlike the bound state
calculation, there is no restriction here on the magnitude of the nonlinear exponent $\alpha$.
For all wavevectors, the transmission increases
with increasing $\alpha$ and does not display any special behavior at $\alpha = 2$.
The increase of $t$ with $\alpha$ can be easily understood with the help of Eq.(\ref{eq:t}):
For any $\alpha >0$, $t^{\alpha} <1$ since $t$ is less than unity. Thus, $\Omega^{*}\ t^{\alpha}
< \Omega^{*}$ which means that the total ``nonlinear'' opacity is always smaller than
the ``linear'' one, hence a higher transmission. 
\vspace{0.2cm}

\noindent
{\bf Summary.}\ \ In this work we have calculated the bound state 
corresponding to a single ``inverted'' nonlinear delta-function potential,
with opacity $\Omega$ and nonlinearity exponent $\alpha$.
Following the usual methods of elementary
quantum mechanics, we arrived at a closed form expression for the
bound state characterized by an exponentially--decreasing
probability profile, with a localization length that decreases
with increasing $\alpha$. The most significant feature of
this solution is the existence of a critical $\alpha$ value,
namely $2$, beyond which the total energy (not the
eigenenergy) of the bound state becomes positive, making
the state unstable against a collapse into the continuum. The
transmission of plane waves across the nonlinear delta potential
is invariant
under a sign change in opacity, and increases
monotonically with an increasing
nonlinearity exponent. The transmission is
always higher than in the linear case,
for a nonzero exponent.

\noindent
Finally, it is important to remark that because of the
nonlinear nature of Eqs. (\ref{eq:1}) and (\ref{eq:total}),
it is no longer possible to superpose stationary states in order
to find the time evolution of a given initial state. A 
stationary state solution of Eq.(\ref{eq:total}) is
now only a particular solution whose relation to
the solution of the time-dependent problem is unclear.
Other features that arise in similar `nonlinear' quantum mechanical
problems include the fact that eigenstates are no longer guaranteed to
be orthogonal to each other. Also, the number of eigenstates
is no longer constant, but depends on nonlinearity.
Thus, `nonlinear' quantum mechanics is considerably more 
challenging than the linear one, although the reader should be
aware that, as was mentioned at the beginning of this Note,
nonlinearity in quantum mechanics is the consequence of some underlying assumption
about the system.

\newpage

\noindent{\bf ACKNOWLEDGMENTS}

\noindent
This work was supported in part by FONDECYT grants 1990960 (M.I.M),
2980033 (C.A.B) and 4990004 (C.A.B). The authors are grateful
to J. R\"{o}ssler for illuminating discussions.

\newpage

\noindent
$1$\hspace{0.4cm}\begin{minipage}[t]{5in}
\baselineskip 24 pt
A. Rabinovitch, ``Negative energy states of an `inverted' delta
potential: Influence of boundary conditions'', Am. J. Phys. {\bf 53},
pp. 768--773 (1985).
\end{minipage}
\vspace{0.3cm}

\noindent
$2$\hspace{0.4cm}\begin{minipage}[t]{5in}
\baselineskip 24 pt
Claude Aslangul, ``$\delta$-well with a reflecting barrier'',
Am. J. Phys. {\bf 63}, pp. 935--940 (1995).
\end{minipage}
\vspace{0.3cm}

\noindent
$3$\hspace{0.4cm}\begin{minipage}[t]{5in}
\baselineskip 24 pt
H. Massmann, ``Illustration of resonances and the law of exponential
decay in a simple quantum-mechanical problem'', 
Am. J. Phys. {\bf 53}, pp. 679--683 (1985); Clinton Dew. Van
Siclen, ``Scattering from an attractive delta-function potential'',
Am. J. Phys. {\bf 56}, pp. 278--280 (1988).
\end{minipage}
\vspace{0.3cm}

\noindent
$4$\hspace{0.4cm}\begin{minipage}[t]{5in}
\baselineskip 24 pt
See for instance, Daniel A. Atkinson, ``An exact treatment of the Dirac delta function potential in the
Schr\"{o}dinger equation'', Am. J. Phys. {\bf 43}, pp. 301--304 (1975);
M. Lieber, ``Quantum mechanics in momentum space: An illustration'',
Am. J. Phys. {\bf 43}, pp. 486--491 (1975); A. K. Jain, S. K. Deb and C. S. Shastry, ``The problem of
several delta-shell potentials in the Lipmann-Schwinger formulation'', Am. J. Phys. {\bf 46}, pp. 147--151 (1978);
Douglas Lessie and Joseph Spadaro, ``One-dimensional multiple scattering in quantum mechanics'',
Am. J. Phys. {\bf 54}, pp. 909--913 (1986);
D. W. L. Sprung, Hua Wu and J. Martorell, ``Scattering by a finite periodic potential'',
Am. J. Phys. {\bf 61}, pp. 1118--1124 (1993);
Indrajit Mitra, Ananda DasGupta and Binayak Dutta-Roy, ``Regularization and renormalization in
scattering from Dirac delta potentials'', Am. J. Phys. {\bf 66}, pp. 1101--1109 (1998). 
\end{minipage}
\vspace{0.3cm}

\noindent
$5$\hspace{0.4cm}\begin{minipage}[t]{5in}
\baselineskip 24 pt
A. Messiah, {\em Quantum Mechanics} (J. Wiley \& Sons, Inc., New York, 1968).
\end{minipage}
\vspace{0.3cm}

\noindent
$6$\hspace{0.4cm}\begin{minipage}[t]{5in}
\baselineskip 24 pt
H. J. Davies and C. S. Adams, ``Mean--field model of a weakly interacting
Bose condensate in a harmonic potential'', Phys. Rev. A {\bf 55}, pp. 2527--2530 (1997).
\end{minipage}
\vspace{0.3cm}

\noindent
$7$\hspace{0.4cm}\begin{minipage}[t]{5in}
\baselineskip 24 pt
T. Holstein, ``Studies of Polaron Motion. Part I. The molecular--crystal model'',
Ann. Phys. (N.Y.) {\bf 8}, 325--342 (1959).
\end{minipage}
\vspace{0.3cm}

\noindent
$8$\hspace{0.4cm}\begin{minipage}[t]{5in}
\baselineskip 24 pt
A. S. Davydov, {\em Theory of molecular excitons} (Plenum Press, New York, 1971);\
A. S. Davydov, {\em Biology and Quantum Mechanics} (Pergamon, Oxford, 1982).
\end{minipage}
\vspace{0.3cm}

\noindent
$9$\hspace{0.4cm}\begin{minipage}[t]{5in}
\baselineskip 24 pt
Peter L. Christiansen and Alwyn C. Scott,
{\em Davydov's Soliton Revisited} (Plenum, New York, 1990).
\end{minipage}
\vspace{0.3cm}

\noindent
$10$\hspace{0.4cm}\begin{minipage}[t]{5in}
\baselineskip 24 pt
J. C. Eilbeck, P. S. Lomdahl and A. C. Scott, ``The Discrete Self--trapping Equation'',
Physica D {\bf 16}, 318--338 (1985).
\end{minipage}
\vspace{0.3cm}

\noindent
$11$\hspace{0.4cm}\begin{minipage}[t]{5in}
\baselineskip 24 pt
P. Yeh, {\em Optical Waves in Layered Media}, New York: Wiley, 1988.
\end{minipage}
\vspace{0.3cm}

\noindent
$12$\hspace{0.4cm}\begin{minipage}[t]{5in}
\baselineskip 24 pt
D. Hennig, G.P. Tsironis, M. Molina and H. Gabriel,
``A Nonlinear Quasiperiodic Kronig-Penney Model'',Phys. Lett. A {\bf 190},
pp. 259--263 (1994); D. Hennig, H. Gabriel, G.P. Tsironis and M. Molina,
``Wave Propagation in Periodic Nonlinear Dielectric Superlattices'',
Appl. Phys. Lett. {\bf 64}, pp. 2934-2936 (1994); N. Sun, D. Hennig, M.I. Molina and
G.P. Tsironis, ``Wave Propagation in a Nonlinear Quasiperiodic Kronig-Penney Lattice'',
J. of Phys: Condens. Matter, {\bf 6}, pp. 7741--7749 (1994); D. Hennig and G. P. Tsironis,
``Wave transmission in nonlinear lattices'', Phys. Rep. {\bf 307}, pp. 333--432 (1999).
\end{minipage}

\newpage
\centerline{{\bf Captions List}}

\vspace{2cm}

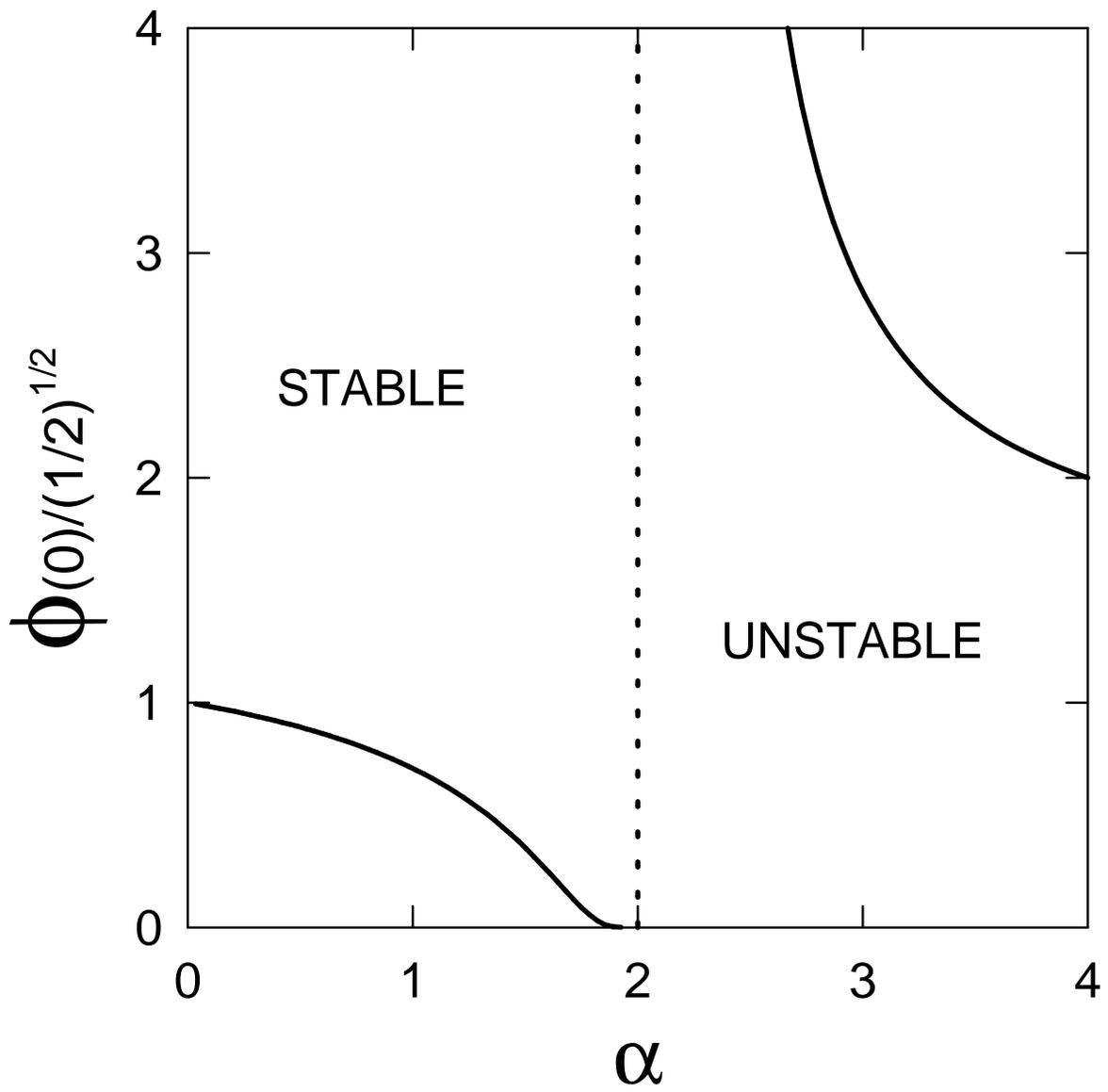
\noindent {\bf Fig.1 :}\ \ Normalized bound state amplitude at the origin as a function of the
nonlinearity exponent. The wavefunction changes discontinuously
at $\alpha = 2$, becoming unstable for $\alpha >2$.
\vspace{0.4cm}

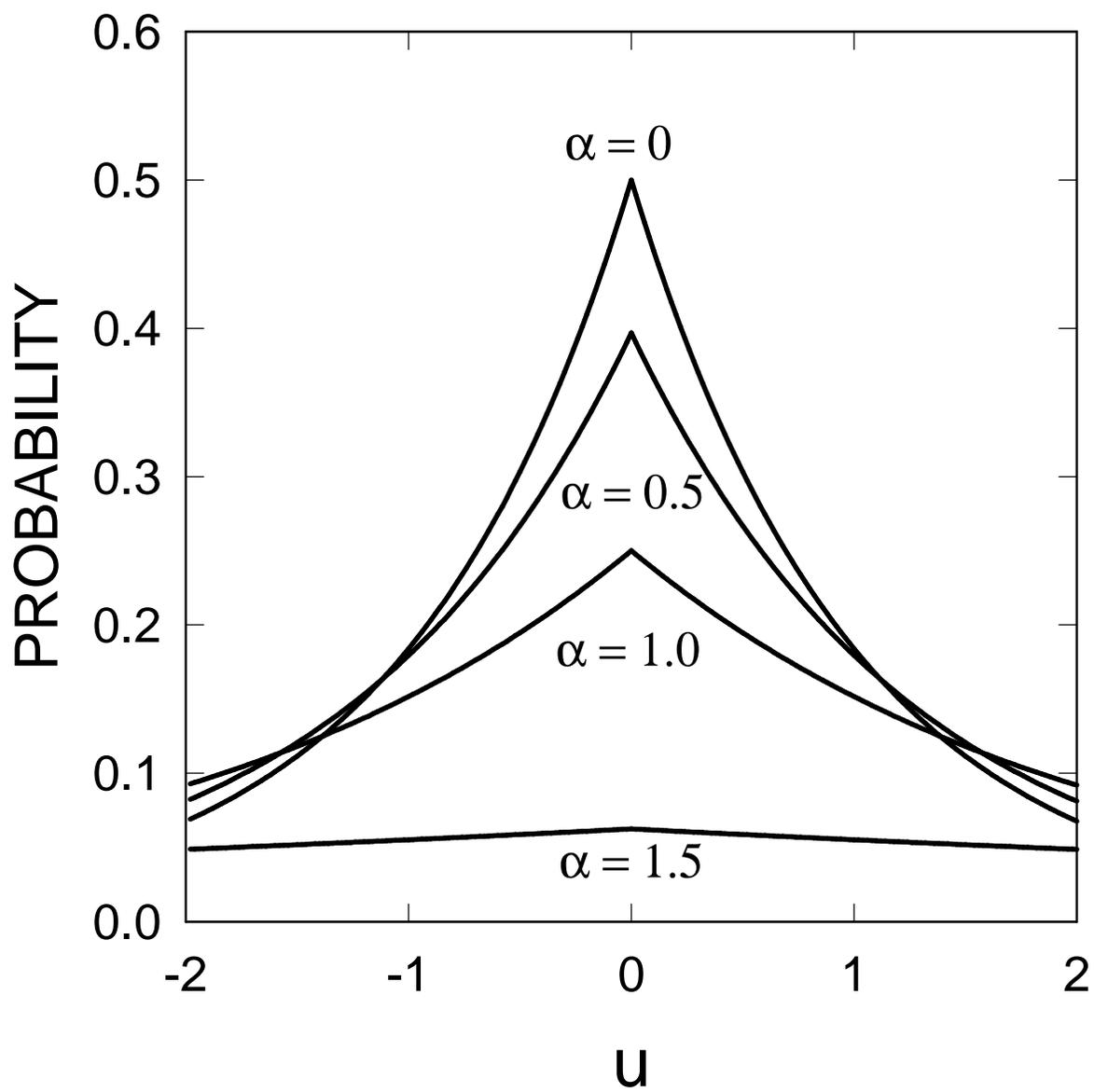
\noindent {\bf Fig.2 :}\ \ Bound state probability profile for the ``inverted'' nonlinear
delta-function potential, for several nonlinearity exponents $\alpha$
that give rise to a stable bound state. 
\vspace{0.4cm}

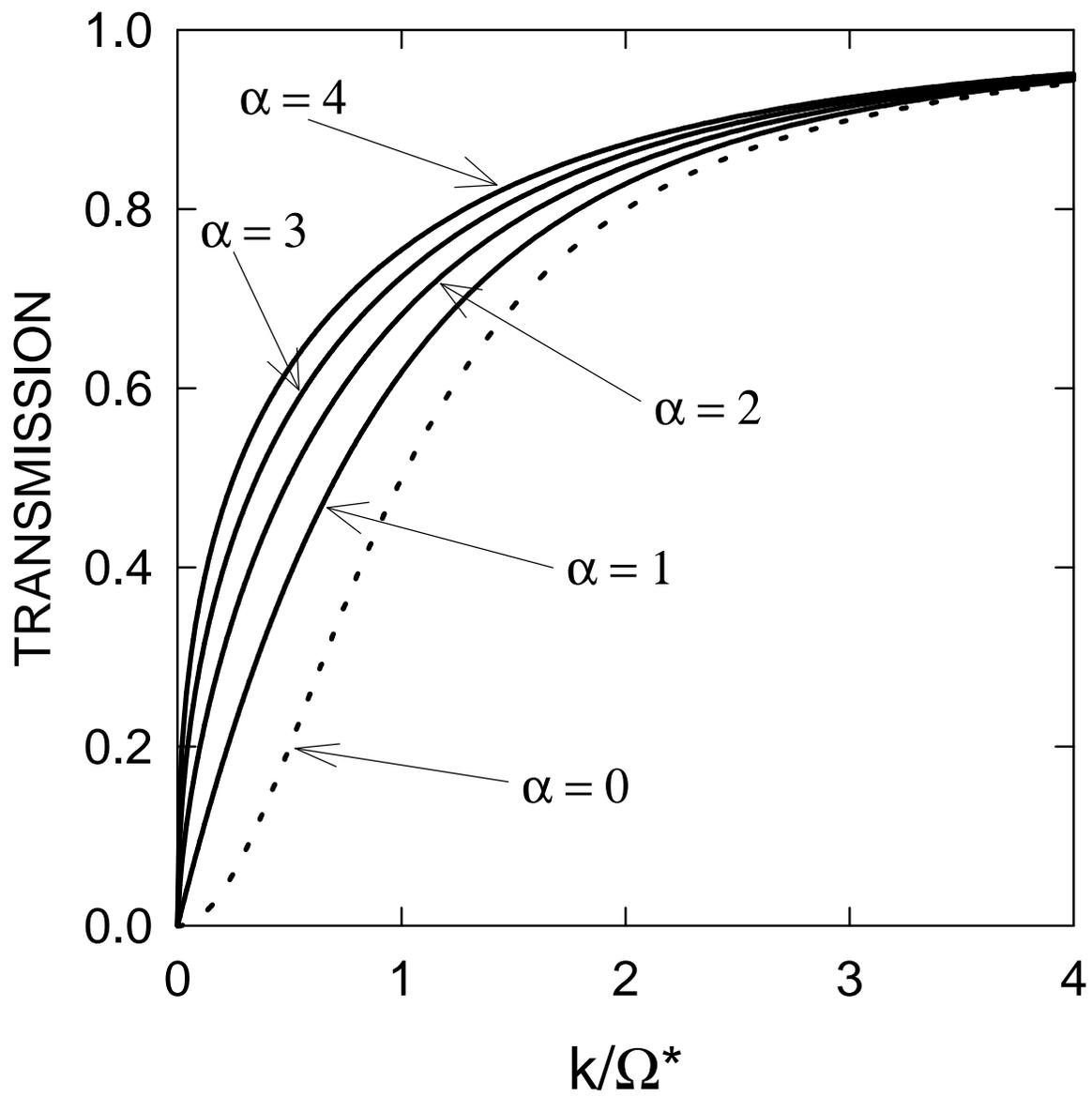
\noindent {\bf Fig.3 :}\ \ Transmission coefficient of plane waves across the nonlinear
delta-function potential versus wavevector, for several different nonlinearity exponent
values. The transmission is the same for the ``upright'' and ``inverted'' delta potentials.

\end{document}